\documentclass[printer,nolinenumbers]{aa} 

\bibpunct{(}{)}{;}{a}{}{,} 

\usepackage{graphicx, color, xcolor}
\usepackage[colorlinks=true,citecolor=blue, linkcolor=blue]{hyperref}%
\usepackage{multirow}
\usepackage{array}
\usepackage{txfonts}

\begin{document}

   \title{Searching for temporary gamma-ray dark blazars associated with IceCube neutrinos}

   \author{E. Kun
          \inst{1,2,3,4,5}
          \and
          I. Bartos\inst{6}
          \and
          J. Becker Tjus\inst{1,3}
          \and
          P. L. Biermann\inst{7,8}
          \and
          A. Franckowiak\inst{2}
          \and
          F. Halzen\inst{9}
          \and
          Gy. Mez\H o\inst{4,5}
          }

   \institute{Theoretical Physics IV: Plasma-Astroparticle Physics, Faculty for Physics \& Astronomy, Ruhr University Bochum, 44780 Bochum, Germany
   \and
Faculty for Physics \& Astronomy, Astronomical Institute, Ruhr University Bochum, 44780 Bochum, Germany
\and
Ruhr Astroparticle And Plasma Physics Center (RAPP Center), Ruhr-Universit\"at Bochum, 44780 Bochum, Germany
\and
Konkoly Observatory, ELKH Research Centre for Astronomy and Earth Sciences, Konkoly Thege Miklós \'ut 15-17, H-1121 Budapest, Hungary
\and
CSFK, MTA Centre of Excellence, Konkoly Thege Miklós \'ut 15-17, H-1121 Budapest, Hungary
\and Department of Physics, University of Florida, PO Box 118440, Gainesville, FL 32611-8440, USA
\and
MPI for Radioastronomy, 53121 Bonn, Germany
\and
Department of Physics \& Astronomy, University of Alabama, Tuscaloosa, AL 35487, USA
\and
Dept. of Physics, University of Wisconsin, Madison, WI 53706, USA
}

   \date{Received ; accepted }

  \abstract
   {Tensions between the diffuse gamma-ray sky observed by the \textit{Fermi} Large Area Telescope (LAT) and the diffuse high-energy neutrino sky detected by the IceCube South Pole Neutrino Observatory question our knowledge about high-energy neutrino sources in the gamma-ray regime. While blazars are among the most energetic persistent particle accelerators in the Universe, studies suggest that they could account for up to for $10$--$30$\% of the 
neutrino flux measured by IceCube.}
   {Our recent results highlighted that the associated IceCube neutrinos arrived in a local gamma-ray minimum (dip) of three strong neutrino point-source candidates. We increase the sample of neutrino-source candidates to study their gamma-ray light curves.}
   {We generate the one-year \textit{Fermi}-LAT light curve for 8 neutrino source candidate blazars (RBS~0958, GB6~J1040+0617, PKS~1313-333, TXS~0506+056, PKS~1454-354, NVSS~J042025-374443, PKS~0426-380 and PKS~1502+106), centered on the detection time of the associated IceCube neutrinos. We apply the Bayesian block algorithm on the light curves to characterize their variability.}
   {Our results indicate that GB6~J1040+0617 was in the phase of high gamma-ray activity, while none of the other $7$ neutrino source candidates were statistically bright during the detection of the corresponding neutrinos and that indeed even most of the times neutrinos arrived in a faint gamma-ray phase of the light curves. This suggests that the $8$ source-candidate blazars (associated with 7 neutrino events) in our reduced sample are either not the sources of the corresponding IceCube neutrinos, or that an in-source effect (e.g. suppression of gamma rays due to high gamma-gamma opacity) complicates the multimessenger scenario of neutrino emission for these blazars.}
   {}


   \maketitle
%

\section{Introduction}
\label{section:intro}

According to observations to date, the energetics of diffuse gamma rays \citep{Fermi2015}, neutrinos \citep{7d5yearsicecubeshowers} and cosmic rays \citep{Auger2017} share similar power-law energetics. At lower energies, though, there is more flux in neutrinos than could be expected from gamma rays, if they are produced by the same sources. Results suggest that there could be a population of neutrino sources being obscured in gamma rays \citep[e.g.][]{7d5yearsicecubeshowers}. The tension at lower energies ($E<100$ TeV) between the diffuse gamma-ray background measured by the \textit{Fermi} Large Area Telescope (\textit{Fermi}-LAT) and the diffuse neutrino background measured by the IceCube Neutrino Observatory also suggests the existence of a population of gamma-ray dark neutrino sources \citep{Murase2016,Murase2019}, for which rather an anti-correlation holds between the neutrino and gamma-ray fluxes, at least in the epochs of efficient neutrino emission. 

After finding a $3\sigma$ significant connection between the gamma-ray flaring blazar TXS~0506+056 and a $\sim290$~TeV IceCube neutrino alert (IC-170922A), this blazar quickly came into the focus of multimessenger astronomy. Follow-up searches in the offline IceCube data revealed a neutrino flare from the direction of TXS~0506+056 at the turn of the year 2014/2015 with an even higher significance ($3.5\sigma$). The individual events of this neutrino flare contained one order of magnitude lower energies ($\sim 10$~TeV) than to the 2017 event, and they do not satisfy the strict selection cuts for single high-energy alerts.

As it turned out, the 2014/2015 neutrino flare \citep{ICTXS2018b} happened during a low gamma-state of the source-candidate blazar TXS~0506+056 \citep{Garrappa2019}. An explanation suggests that if the medium is ideal to generate neutrinos it will absorb the pionic gamma photons that are also generated in the hadronic process \citep{Halzen2019,Halzen2020}. Observations suggest this might be actually true, at least for a population of high-energy (HE) neutrino sources \citep[e.g.][]{Kun2021}. The PeV--neutrino event called Big Bird arrived in the local minimum of a $\sim 85$-days wide dip of the source candidate blazar PKS~B1424-418 \citep{Kadler2016}. The neutrino event IC-190730A was detected from the direction of PKS~1502+106 when this blazar was in the middle of a deep, about $180$ days wide gamma-ray minimum \citep{Kun2021}. The neutrino event IC-170922A was observed in a $\sim 28$ days wide gamma-ray dip of TXS~0506+056. 

As was discussed in \citet{Kun2021}, the temporal gamma-suppression potentially resolves the apparent contradiction of the blazar models simultaneously producing a detectable neutrino flux and a gamma flare, since at the time of efficient neutrino production the observed gamma photon flux drops, because the energy of pionic gamma rays cascade down to X-ray energies. 

In this paper we examine $8$ \textit{Fermi}-LAT blazars with good photon statistics \citep[][]{Kun2022} which we found within the 90\% containment area of the $70$ IceCube track-type neutrino events. In Section \ref{section:sourceselection}. we introduce our previous search to find neutrino source candidates in the Fermi-LAT 4FGL-DR2 catalog, and summarize source properties of individual neutrino source candidates. In Section \ref{section:gammaanalysis}. we present our \textit{Fermi}-LAT light curve analysis. In Section \ref{section:results}. we present our results, while in Section \ref{section:disc}. we discuss them and give our conclusions.

\section{Source selection}
\label{section:sourceselection}
\subsection{Previous catalog search to find \textit{Fermi}-LAT 4FGL-DR2 neutrino source candidates}


In \citet{Kun2022}, we searched gamma-ray, X-ray and radio point sources in the 90\% containment area of the $70$ IceCube track-type neutrino events recorded between 2009 October 13 and 2019 September 30 and listed by \citet{Giommi2020}. We found 29 gamma-ray neutrino source candidates in the Fermi-LAT 4FGL-DR2 catalog \citep{Fermi4FGL2020}, 61 such X-ray point sources in the Swift-XRT Point Source catalog \citep[keeping only the AGN-type objects,][]{Evans2020} and 87(96) neutrino counterparts in the Combined Radio All-Sky Targeted Eight GHz Survey \citep{Healey2007} at 4.8 GHz (8.4 GHz). These are the so-called $\nu$-samples.

After the construction of the $\nu$-samples, we made a randomness test of them to see if our selection method chooses random $\nu$-samples or $\nu$-samples with flux properties being significantly different from the full catalogs. We note that we did not scramble the declinations of the neutrinos because of the strong variation of the sensitivity of IceCube with declination. We found that the constructed $\nu$-samples represent special samples of sources taken from the full \textit{Fermi}-LAT 4FGL-DR2/\textit{Swift}-XRT 2SXPS/CRATES catalogs with similar significance ($2.1\sigma$, $1.2\sigma$, $2\sigma$ at $4.8~\mathrm{GHz}$, $2.1\sigma$ at $8.4~\mathrm{GHz}$, respectively). In this sense special means the neutrinos do not randomly select samples. We already analyzed the neutrino-radio connection in \citet{Kun2022}. After collecting redshifts and deriving sub-samples of the CRATES catalog complete in the redshift--luminosity plane, we found that the 4.8 GHz ($8.4$~GHz) sub-sample can explain (90\% C.L.) between 4\% and 53\% ($3$\% and $42$\%) of the $70$ IceCube track-type neutrino events listed by \citet{Giommi2020}.

In this paper we focus on the possible neutrino-gamma ray connection. We list the name, the sky coordinates, the flux properties, the class and the association of the 29 neutrino source candidate \textit{Fermi}-LAT blazars in Table \ref{table:4fglfindinfg}. We will generate the gamma-ray light curves plus minus half year around the detection time of the associated neutrinos. For sources with integrated flux below $10^{-9}\mathrm{ph~cm}^{-2}~\mathrm{s}^{-1}$ in the energy range from $0.1$ to $100$ GeV, and not experiencing bright flaring activity in the time window, there is not enough photon statistics to obtain good resolution on the light curve. Therefore we apply the above gamma-ray photon flux threshold, and obtain a sample of 8 candidate sources (associated with 7 neutrino events) from this selection that we will study in the followings. In the next subsection we introduce the 8 neutrino source candidate \textit{Fermi}-LAT blazars which constitute our sample of interest in this study.

\begin{table*}
\caption{Source properties of the $29$ neutrino source candidates  from the $10$ years \textit{Fermi}-LAT point--source catalog (4FGL-DR2), published in \citet{Kun2022}. The columns are: (1) 4FGL source ID, (2) right-ascension (J2000, in degrees), (3) declination (J2000, in degrees), (4) predicted number of photons, (5) photon flux between $0.1$--$100$~GeV, (6) peak photon flux, (7) class: BLL=BL Lac class of blazar, FSRQ=FSRQ class of blazar, BCU=active galaxy of uncertain class, (8) association name. Sources with $F_{1000}$ values above $10^{-9} ~\mathrm{ph~cm}^{-2}~\mathrm{s}^{-1}$ are marked by boldface.  $^\star$: this source is within 1.5 times the 90\% containment area of the respective neutrino and was included in error in \citet{Kun2022}. }
\label{table:4fglfindinfg}
\centering
\begin{tabular}{cccccccc}
\hline\hline
{4FGL~ID} & {RA} & {Dec} & {N$_\mathrm{pred}$} & {$F_{1000}$} & {$F_\mathrm{peak}$} & {Class} & {Association}\\
 & ($^\circ$) & ($^\circ$) & & ($\mathrm{ph~cm}^{-2}~\mathrm{s}^{-1}$) & ($\mathrm{ph~cm}^{-2}~\mathrm{s}^{-1}$) &  & \\
(1) & (2) & (3) & (4) & (5) & (6) & (7) & (8)\\
\hline
4FGL~J2030.5+2235 & $307.64$ & $22.59$ & $117.0$ & 1.36e-10 & -- & -- &-- \\
4FGL~J2030.9+1935 & $307.74$ & $19.60$ & $526.8$ & 8.24e-10 & -- & BLL & RX~J2030.8+1935\\
4FGL~J1808.2+3500 & $272.07$ & $35.01$ & $730.8$ & 3.72e-10 & 1.75e-08 & BLL & MG2~J180813+3501\\
4FGL~J1808.8+3522 & $272.22$ & $35.38$ & $305.6$ & 1.91e-10 & -- & BLL & 2MASX~J18084968+3520426\\
4FGL~J1744.2-0353 & $266.05$ & $-3.89$ & $740.6$ & 3.90e-10 & -- & FSRQ & PKS~1741-03\\
4FGL~J0230.3+1713 & $37.60$ & $17.22$ & $537.5$ & 2.52e-10 & -- & -- &--\\
4FGL~J0224.9+1843 & $36.23$ & $18.72$ & $983.7$ & 1.70e-10 & 4.52e-08 & FSRQ & TXS~0222+185\\
4FGL~J1117.0+2013 & $169.27$ & $20.23$ & $1442.3$ & \textbf{1.25e-09} & 2.09e-08 & BLL & RBS~0958\\
4FGL~J2227.9+0036 & $336.98$ & $0.62$ & $679.9$ & 8.70e-10 & -- & BLL & PMN~J2227+0037\\
4FGL~J1233.0+1333 & $188.26$ & $13.56$ & $828.0$ & 3.34e-10 & -- & -- &--\\
4FGL~J1040.5+0617 & $160.15$ & $6.28$ & $2927.1$ & \textbf{1.49e-09} & 6.26e-08 & BLL & GB6~J1040+0617\\
4FGL~J0854.0+2753 & $133.52$ & $27.88$ & $28.4$ & 3.93e-11 & -- & BLL & SDSS~J085410.16+275421.7\\
4FGL~J1557.9-0001 & $239.49$ & $-0.02$ & $412.3$ & 1.86e-10 & -- & FSRQ & PKS~1555+001\\
4FGL~J1258.7-0452 & $194.68$ & $-4.87$ & $124.1$ & 1.73e-10 & -- & BLL & RBS~1194\\
4FGL~J1311.8+2057 & $197.97$ & $20.96$ & $617.2$ & 5.30e-11 & -- & BCU & MG2~J131144+2052\\
4FGL~J0103.5+1526 & $15.88$ & $15.43$ & $294.7$ & 1.74e-10 & -- & BLL & TXS~0100+151\\
4FGL~J1316.1-3338 & $199.03$ & $-33.64$ & $3866.4$ & \textbf{2.39e-09} & 1.22e-07 & FSRQ & PKS~1313-333\\
4FGL~J0244.7+1316 & $41.19$ & $13.28$ & $574.5$ & 2.01e-10 & 1.78e-08 & Blazar & GB6~J0244+1320\\
4FGL~J1447.0-2657 & $221.77$ & $-26.96$ & $195.1$ & 1.64e-10 & -- & BCU & NVSS~J144657-265713\\
4FGL~J1439.5-2525 & $219.88$ & $-25.42$ & $153.9$ & 1.55e-10 & -- & BCU & NVSS~J143934-252458\\
4FGL~J0509.4+0542 & $77.36$ & $5.70$ & $7619.8$ & \textbf{8.02e-09} & 1.94e-07 & BLL & TXS~0506+056\\
4FGL~J1457.4-3539 & $224.37$ & $-35.65$ & $5233.2$ & \textbf{3.49e-09} & 2.56e-07 & FSRQ & PKS~1454-354\\
4FGL~J1505.0-3433 & $226.26$ & $-34.55$ & $642.8$ & 3.77e-10 & 2.15e-08 & BLL & PMN~J1505-3432\\
4FGL~J2351.4-2818 & $357.87$ & $-28.31$ & $290.7$ & 1.28e-10 & -- & --& --\\
4FGL~J0420.3-3745 & $65.09$ & $-37.75$ & $1841.0$ & \textbf{1.09e-09} & 3.43e-08 & BCU & NVSS~J042025-374443\\
4FGL~J0428.6-3756 & $67.17$ & $-37.94$ & 24240.3 & \textbf{2.36e-08} & 3.57e-07 & BLL &  PKS~0426-380\\
4FGL~J1504.4+1029 & $226.10$ & $10.50$ & $25352.5$ & \textbf{2.02e-08} & 9.72e-07 & FSRQ & PKS~1502+106\\
4FGL~J0946.2+0104 & $146.57$ & $1.07$ & $201.8$ & 2.15e-10 & -- & BLL & 1RXS~J094620.5+010459\\
4FGL~J0948.9+0022 & $147.24$ & $0.37$ & $7769.2$ & \textbf{2.29e-09} & 1.61e-07 &  NLSY1 & PMN~J0948+0022$^\star$\\
\hline
\end{tabular}
\end{table*}
\begin{table*}
\caption{Source properties of neutrino source candidates from 4FGL-DR2 being considerable bright in our observational window of one year ($F_{1000}>10^{-9} ~\mathrm{ph~cm}^{-2}\,\mathrm{s}^{-1}$, see with boldface in Table \ref{table:4fglfindinfg}). The columns are: (1) source name, (2) 4FGL source ID, (3) right-ascension (J2000, in degrees), (4) declination (J2000, in degrees), (5) type, (6) redshift, (7) neutrino ID (8), neutrino right-ascension (J2000), (9) neutrino declination (J2000), (10) time window for \textit{Fermi}-LAT analysis.\label{table:9sources}}
\centering
\resizebox{\textwidth}{!}{\setlength{\extrarowheight}{4pt}{\begin{tabular}{cccccccccc}
\hline\hline
Source & 4FGL ID & RA & Dec & Class & $z$ & {ID$_\nu$} & RA$_\nu$ & DEC$_\nu$ & $T_\mathrm{MJD,\gamma}$\\
 &  & ($^\circ$) & ($^\circ$) &  &  &  & ($^\circ$) & ($^\circ$) & (days)\\
 (1) & (2) & (3) & (4) & (5) & (6) & (7) & (8) & (9) & (10)\\
\hline
RBS~0958 & J1117.0+2013 & $169.27$ & $20.23$ & BLL & $0.1379$ & IC-130408A & $168.16^{+2.87}_{-1.9}$ & $20.67^{+1.15}_{-0.89}$ & [$56207.6$--$56572.8$]\\
GB6~J1040+0617 & J1040.5+0617 & $160.15$ & $6.28$ & BLL & $0.7351$ & IC-141209A & $159.81^{+0.84}_{-1.04}$ & $6.57^{+0.64}_{-0.56}$ & [$56817.5$--$57182.8$]\\
PKS~1313-333 & J1316.1-3338 & $199.03$ & $-33.64$ & FSRQ & $1.210$ & IC-160814A & $199.39^{+2.43}_{-3.03}$ & $-32.4^{+1.39}_{-1.21}$ & [$57432.3$--$57797.5$]\\
TXS~0506+056 & J0509.4+0542 & $77.36$ & $5.70$ & BLL & $0.3365$ & IC-170922A & $77.43^{+0.95}_{-0.65}$ & $5.72^{+0.5}_{-0.3}$ & [$57836.2$--$58201.5$]\\
PKS~1454-354 & J1457.4-3539 & $224.37$ & $-35.65$ & FSRQ & $1.424$ & IC-181014A & $224.3^{+1.4}_{-2.85}$ & $-34.8^{+1.15}_{-1.85}$ & [$58222.9$--$58588.1$]\\
NVSS~J042025-374443 & J0420.3-3745 & $65.09$ & $-37.75$ & BCU & $0.300$ & IC-190504A & $65.79^{+1.23}_{-1.23}$ & $-37.44^{+1.23}_{-1.23}$ & [$58425.1$--$58790.4$]\\
PKS~0426-380 & J0428.6-3756 & $67.17$ & $-37.94$ & FSRQ & $1.111$ & IC-190504A & $65.79^{+1.23}_{-1.23}$ & $-37.44^{+1.23}_{-1.23}$ & [$58425.1$--$58790.4$]\\
PKS~1502+106 & J1504.4+1029 & $226.10$ & $10.50$ & FSRQ & $1.839$ & IC-190730A & $225.52^{+1.28}_{-1.43}$ & $10.47^{+1.14}_{-0.89}$ & [$58512.2$--$58877.5$]\\
\hline
\end{tabular}}}
\end{table*}

\subsection{Individual neutrino source candidates}

\textbf{RBS~0958} is a BL Lac object at redshift $z=0.13793\pm0.00004$ \citep{Ahn2012}, associated with the \textit{Fermi}-LAT source 4FGL~J1117.0+2013. RBS 0958 is a high-synchrotron peaked \citep[e.g.][]{Paiano2017} VHE candidate \citep{Furniss2015}, and is considered to be a candidate TeV emitter based on the position of its synchrotron peak \citep{Costamante2018}. RBS 0958 is observed by the High Energy Stereoscopic System (H.E.S.S.), with an integral flux upper limit of $I(>610~\mathrm{GeV})=1.44\times 10^{-12}~\mathrm{ph~cm}^{-2}~s^{-1}$ \citep{Aharonian2015}.

\textbf{GB6 J1040+0617} is a low-synchrotron peaked BL Lac object at a redshift $z=0.7351\pm0.0045$ \citep{Ahn2012,Maselli2015}, associated with the \textit{Fermi}-LAT source 4FGL J1040.5+0617. \citet{Garrappa2019} highlighted the increased gamma-ray activity of this source for a period of $\sim90$ days around the time of the neutrino detection. \citet{Gabanyi2019} analyzed $3.4 \mu m$ and $4.6\mu m$ band GB6 J1040+0617 data of the Wide-field Infrared Survey Explorer (\textit{WISE}). They found the source to be the brightest, increasing its flux density by almost $90$\%, in the mission phase preceding the neutrino event IceCube-141209A by $4.5$ days. 

\textbf{PKS 1313-333} is a flat sepctrum radio quasar (FSRQ) at a redshift $z\approx1.210$ \citep[][]{Jauncey1982}, associated with the \textit{Fermi}-LAT source 4FGL J1316.1-3338. \textit{Fermi}-LAT detected an increasing gamma-ray flux from a source positionally consistent with PKS 1313-333\footnote{\url{https://www.astronomerstelegram.org/?read=8533}}, that was confirmed by Astro-rivelatore Gamma a Immagini Leggero (AGILE)\footnote{\url{https://www.astronomerstelegram.org/?read=8536}}. Integrating photons from 2016-01-10 02:00 UT to 2016-01-12 02:00 UT, a maximum likelihood analysis of AGILE data yields the detection of the source at a significance level of about $5\sigma$ with a flux of $(1.6 \pm 0.6) \times 10^{-6}~\mathrm{ph~cm}^{-2}~\mathrm{s}^{-1} (E > 100~\mathrm{MeV})$, in agreement with the \textit{Fermi}-LAT measurement. An increasing gamma-ray flux was observed again in 2021\footnote{\url{https://www.astronomerstelegram.org/?read=14672}}.

\textbf{TXS~0506+056} is a BL Lac object \citep[suggested to have at least temporarily intrinsic properties of a FSRQ,][]{Padovani2019} at a redshift $z = 0.3365 \pm 0.0010$ \citep{Paiano2018}, associated with the \textit{Fermi}-LAT source 4FGL~J0509.4+0542. This source is a TeV\footnote{\url{http://tevcat.uchicago.edu/?mode=1&showsrc=309}} emitter. A multimessenger campaign, triggered by the coincident observation of a gamma-ray flare and the $290$~TeV IceCube neutrino IC-170922A (assuming $\varepsilon_\nu^{-2.19}$ neutrino spectrum), identified TXS 0506+056 as the possible source of high-energy neutrinos at a $3\sigma$ significance level \citep{ICTXS2018a}. With this, TXS 0506+056 became the first identified source of high-energy cosmic neutrinos at such high significance. The blazar TXS 0506+056 is $0.1\degr$ away from best-fit neutrino location of IC-170922A (56.5\% signalness). The quantity signalness is the probability that an observed neutrino has an astrophysical ($s =
1$) or an atmospheric ($s = 0$) origin \citep[see its definition in][]{Aartsen2017realtime}. Going back to the archival data, the IceCube Collaboration found a 160-day long neutrino flare in 2014/2015 from the direction of TXS 0506+056 at a significance level of $3.5\sigma$ \citep{ICTXS2018b}. Despite being more dominant than IC-170922A, this neutrino flare happened during a low-gamma state of TXS 0506+056 \citep{Garrappa2019}. Putting this behavior into a common physical picture with IC-170922A and the coincident gamma-flare still challenges the current blazar models.

The Major Atmospheric Gamma Imaging Cherenkov Telescopes (MAGIC) measured the upper limit on the $>90$ GeV flux of TXS 0506+056 as $<3.56\times 10^{-11}~\mathrm{ph~cm}^{-2}~\mathrm{s}^{-1}$ in the epoch MJD~$58020.18$ \citep[][]{Ansoldi2018}, about 1 day after the detection of IC-170922A, and MAGIC measured a significant increase in $>90$ GeV flux,  $(4.7\pm1.4)\times 10^{-11}~\mathrm{ph}\,\mathrm{cm}^{-2}\,\mathrm{s}^{-1}$ at MJD~$58029.22$, and $(8.7\pm2.0)\times 10^{-11}~\mathrm{ph}\,\mathrm{cm}^{-2}\,\mathrm{s}^{-1}$ at MJD~$58030.24$, 10--11 days after the detection of IC-170922A. Confirming this elevated high-energy state, the Very Energetic Radiation Imaging Telescope Array System (VERITAS) Collaboration reported the integral flux ($>110$ GeV) of TXS 0506+056 as $(8.9\pm 1.6) \times 10^{-12}~\mathrm{ph}\,\mathrm{cm}^{-2}\, \mathrm{s}^{-1}$ \citep{Abeysekara2018}. H.E.S.S. performed follow-up observations of the high-energy neutrino on 23 September 2017 and the following night (24 September 2017 at
03:10 UTC). A preliminary on-site calibration and analysis searching for a point-like gamma-ray source from within the 90\% uncertainty region of the neutrino event IC-170922A revealed no significant detection \citep[upper limit $< 1.0\times 10^{-11}~\mathrm{ph}~\mathrm{cm}^{-2}~\mathrm{s}^{-1}$ for $>175$ GeV,][]{Hess2019}. Employing the ANTARES neutrino telescope data recorded in a cone of $3\degr$ centered on the best-fit coordinates of IC-170922A and within a $\pm1$ h time-window centered on the event time, \citet{Albert2018} constrained the neutrino fluence as $<15~\mathrm{GeV}~\mathrm{cm}^{-2}$ for $\gamma=2$ (integrated between $3.3$~TeV and $3.4$~PeV) and as $<34~\mathrm{GeV}~\mathrm{cm}^{-2}$ for $\gamma=2.5$ (integrated between $450$~GeV and $280$~TeV) assuming a point source with power-law spectrum $dN/dE\sim E^{-\gamma}$.

The radio flux density of TXS~0506+056 measured by the Owens Valley 40m Radio telescope started to increase in the beginning of 2016, and was still increasing after the neutrino detection. During the 2014/2015 neutrino flare the radio flux density was average. 
With very long baseline interferometry (VLBI), the radio jet of TXS 0506+056 shows strong signs of spine-sheath structure within the inner 1 mas, corresponding to about 70-140 pc in de-projected distance, from the millimeter-VLBI core \citep{Ros2020txs}. It has been shown, that at 15 GHz \citep[e.g.][]{Kun2019} and at 43 GHz \citep{Ros2020txs} the VLBI core is responsible for the radio brightening. Several scenarios have been suggested to explain the neutrino emission \citep[][]{Britzen2019,Garrappa2019,Reimer2019,Petropoulou2020,Zhang2020}.

\textbf{PKS 1454-354} is a low-synchrotron peaked FSRQ type blazar at redshift $z\approx1.424$ \citep{Jackson2002}, associated with the \textit{Fermi}-LAT source 4FGL J1457.4-3539. This source is located within the 90\% error ellipse of the neutrino IC-181014A. The \textit{Fermi}-LAT collaboration reported an increased gamma-ray activity from the source in September of 2008 \citep[][]{Abdo2009a}. PKS 1454-354 has a weak one-sided jet. 

\textbf{NVSS J042025-374443} is a radio source at the (photometric) redshift $z\approx0.3$ \citep{Malkin2018}, associated with the \textit{Fermi}-LAT source 4FGL J0420.3-3745. This is another $>100$ MeV gamma-ray source located within the 90\% IC-190504A localization error, at a distance of roughly 37 arcmin away from the best-fit neutrino localization (PKS 0426-380 is located at a distance of roughly 72 arcmin).

\textbf{PKS 0426-380} is a reclassified FSRQ at redshift $z\approx1.111$ \citep{Heidt2004,Sbarufatti2005}, associated with the \textit{Fermi}-LAT source 4FGL J0428.6-3756. This source is located within the 90\% error ellipse of the neutrino IC-190504A. \citet{Tanaka2013} reported the \textit{Fermi}-LAT detection of two very-high-energy ($E \sim 100$ GeV) gamma-ray photons from the directional vicinity of PKS 0426-380. \citet{Zhang2017b} found a possible gamma-ray quasi-periodicity of this blazar with a period of $3.35 \pm 0.68$ yr at the significance level of approximately $3.6\sigma$. 
 
\textbf{PKS 1502+106} is an FSRQ, the 15th brightest gamma-ray source at $>100$ MeV in terms of energy flux among $2,863$ sources in the fourth catalog of AGN detected by \textit{Fermi}-LAT \citep[4LAC][]{Ajello2020}. The redshift of PKS 1502+106 was estimated to be $z=1.839$ \citep[][]{Smith1977}, later confirmed by the good signal-to-noise ratio spectrum of the Sloan Digital Sky Survey \citep[SDSS; $z = 1.8385 \pm 0.0024$ at high confidence,][]{Schneider2010A}. Given its distance and still enormous observed gamma-ray photon flux, PKS 1502+106 must have extremely high intrinsic luminosity. This source is highly variable in the \textit{Fermi}-LAT gamma-ray band \citep{Abdo2010a}. The source showed strong gamma-ray flares in the \textit{Fermi}-LAT energy range in 2008 \citep[][]{Abdo2010b}. The flaring activity of PKS 1502+106 was renewed between 2009 and 2010. From 2015, the source showed more, strong gamma-ray flares \cite[see e.g. in][]{Kun2021}.

VLBI observations revealed a one-sided, curved radio jet+bright core structure at pc-scales \citep[e.g.][]{Karamanavis2016}. At radio frequencies, the source is also highly variable. 

The IceCube Detector recorded a high-energy neutrino event on June 30, 2019 (with 67\% signalness), that was distributed through the Gold alert stream \citep{2019GCN.25225....1I}. The most probable neutrino energy was approximately $300$ TeV, assuming a $\varepsilon_\nu^{-2.19}$ neutrino spectrum. PKS 1502+106 lies with an offset of $0.31\degr$ from the best-fit neutrino location of IC-190730A. At the time of the corresponding neutrino detection, PKS 1502+106 was in quiet state the in optical \citep{2019ATel12974....1S,2019ATel12971....1L}, UV \citep{2020ApJ...893..162F}, X-ray \citep{2019ATel12985....1S,Kun2021}, and gamma-ray regimes \citep{2019GCN.25239....1G}. A long-term radio outburst that started in 2014 had just reached its peak flux when the IceCube neutrino was recorded \citep{2019ATel12996....1K}. This source became part of the Icecube and ANTARES a priori defined monitored source list (34 gamma-ray bright AGN). 

After the detection of IC-190730A, the nearby candidate PKS 1502+106 quickly became the center of focus of many studies. \citet{Kun2021} studied cosmic neutrinos from temporarily gamma-suppressed blazars, such as PKS 1502+106. As it was shown, the neutrino event IC-190730A \citep{IceCube190730A} arrived during an exceptionally low gamma-ray state of PKS 1502+106, that lasted for several months. \citet{Rodrigues2021} identified one quiescent state and two flaring states with hard and soft gamma-ray spectra. They found two hadronic models that can describe the multiwavelength emission during all three states: a leptohadronic model with a contribution from photohadronic processes to X-rays and gamma rays, and a proton synchrotron model, where the emission from keV to $10$ GeV comes from proton synchrotron radiation. \citet{Oikonomou2021} analyzed UV, optical, X-ray observations of PKS 1502+106, as well as collecting other multiwavelength data. They found that the leptohadronic modeling, in which the multiwavelength emission of PKS 1502+106 originates beyond the broad-line region (BLR) and inside the dust torus, is most consistent with the observations. Analyzing $15$ GHz Very Long Baseline Array (VLBA) and astrometric 8 GHz VLBA data, \citet{Britzen2021} presented evidence for a radio ring structure that develops with time. They suggested the neutrino IC-190730A was most likely produced by proton-proton interaction in the blazar zone beyond the BLR, enabled by episodic encounters of the jet with dense clouds, i.e. some molecular clouds in the narrow line region (NLR).

\section{Analysis of \textit{Fermi}-LAT data of 8 blazars}
\label{section:gammaanalysis}

The LAT instrument onboard the \textit{Fermi} Gamma-ray Space Telescope pair conversion gamma-ray telescope is designed to cover the energy band from 20 MeV to greater than 300 GeV. In this study we employed one year of Pass8 \textit{Fermi}-LAT\footnote{\url{https://fermi.gsfc.nasa.gov/science/instruments/lat.html}} data\footnote{\url{https://fermi.gsfc.nasa.gov/ssc/data/analysis/documentation/Pass8_usage.html}} for each of the eight \textit{Fermi}-LAT neutrino source candidates, in a time range plus minus a half year centered on the detection time of the corresponding neutrinos as listed in Table \ref{table:9sources}. The region of interest (ROI) was centered at the J2000 sky coordinates of the 4FGL-DR2 sources encompassing an area of the sky within $15\degr$ ROI radius. We selected event type "front+back" (evtype=3) which is the recommended type for a point source analysis.

We generated the likelihood light curve of the eight brightest 4FGL-DR2 neutrino source candidates. We performed a binned likelihood analysis of the data utilizing the fermipy v1.0.1 and ScienceTools v2.0.8 packages, both built in the FermiBottle Docker container and analysis environment provided by the Fermi Science Support Center (FSSC)\footnote{\url{https://github.com/fermi-lat/FermiBottle}}. The analysis was carried out on a cluster of $16$ vCPU-s (Intel Skylake $16 \times 2.2$ GHz) of the ELKH Cloud\footnote{\url{https://science.cloud.hu/}}. The instrument response functions \verb+P8R3_SOURCE_V2+ were employed together with templates of the Galactic interstellar emission model \verb+gll_iem_v07.fits+ and of the isotropic diffuse emission \verb+iso_P8R3_SOURCE_V2_v1.txt+\footnote{\url{https://fermi.gsfc.nasa.gov/ssc/data/access/lat/BackgroundModels.html}}. We applied the nominal data quality cut \verb+(DATA_QUAL > 0) && (LAT_CONFIG==1)+, and a zenith angle cut $\theta<90\degr$ to eliminate Earth limb events. Time intervals when the Sun was closer than $15\degr$ to the target sources were filtered out. 

The detection level of each source was measured by the definition of test statistic \citep[$TS$,][]{Mattox1996}, and the detection limit of new sources was set to $TS_\mathrm{min}=25$ ($\sim 5 \sigma$). The test statistic is defined as $TS=-2\mathrm{ln}(L_\mathrm{max,1}/L_\mathrm{max,0})$, where $L_\mathrm{max,0}$ is the maximum likelihood value for a model without an additional source (the 'null hypothesis') and $L_\mathrm{max,1}$ is the maximum likelihood value for a model with the additional source at a specified location.
 
Since the two models are nested, according to Wilks' theorem \citep{Wilks1938} the $TS$--distribution will asymptotically follow a $\chi^2$ distribution. Then the TS distribution should be drawn from a ${\chi^2}_n$ distribution, where $n$ is the difference in the degree of freedom between the models with and without the additional source. In our \textit{Fermi}-LAT analysis the normalization factor of the Galactic and isotropic diffuse components and sources with $TS>25$ were freed. 
We associated power-law profiles to the additional sources that are detected in our analysis, but do not appear in the 4FGL-DR2. We consider new point sources to be separate sources if the angular distance between them is larger then $0.3$ degrees. We present a summary of additional sources found in the ROI of $7$ target sources in Table \ref{table:addsources} of the Appendix. To get more information about variability in the light curves we applied the adaptive-binning algorithm \citep[][]{Lott2012}, setting the relative flux uncertainty to be 15\%. With adaptive binning, our analysis is more sensitive to the light variations. To achieve the same photon statistics throughout the light curve, the bins are shorter when the source is bright and the bins are wider when the source is fainter.

\begin{figure*}
\centering
\includegraphics[width=0.4\textwidth, angle=270]{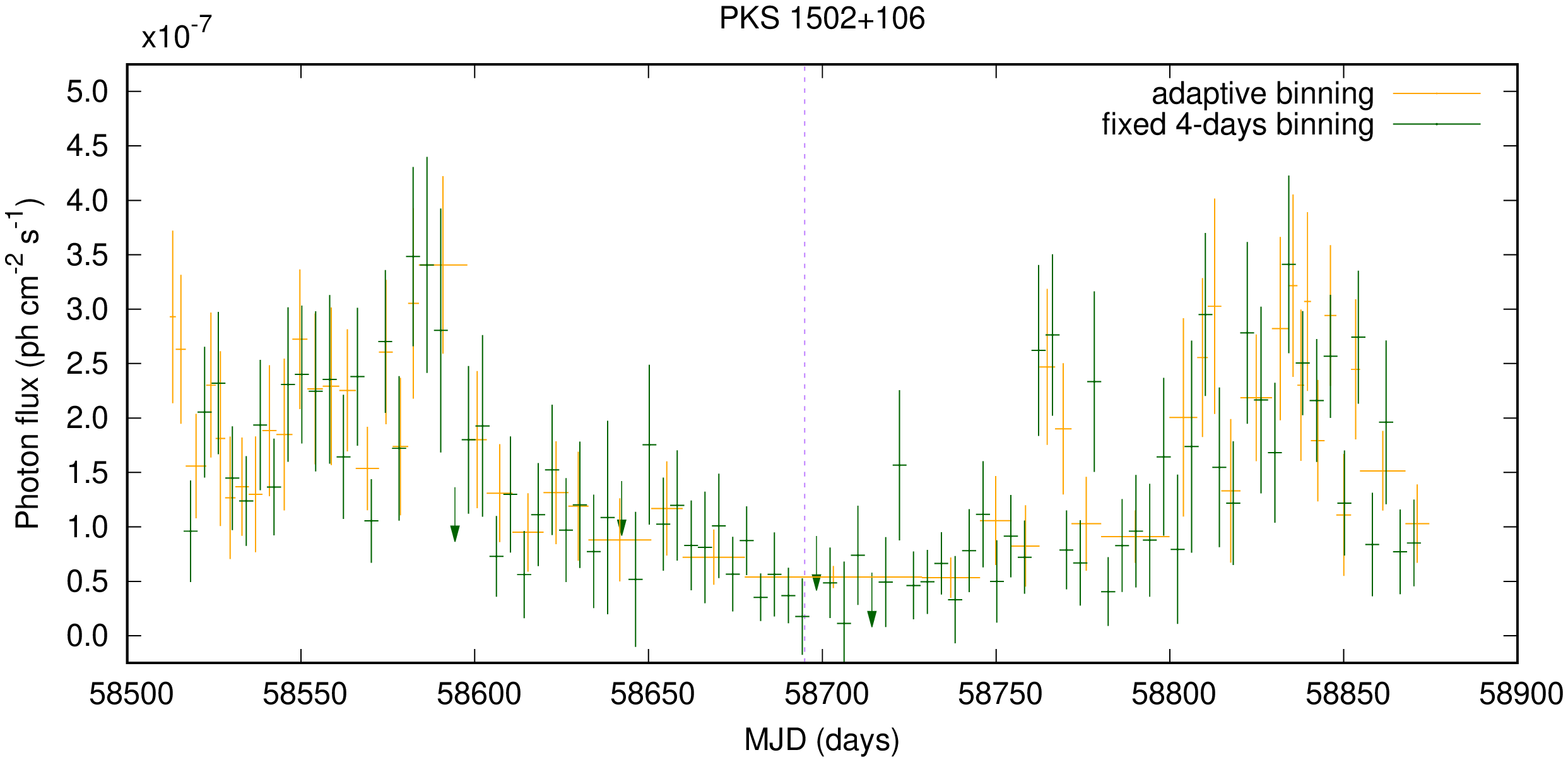}
\caption{This plot shows the light curve of PKS~1502+106 with fixed binning (green crosses), overplotted on its adaptively-binned light curve (orange crosses). For very faint bins of the fixed binning light curve we plot only upper limits (down-pointing green arrows).}
\label{fig:comp_adapt_fixed_pks}
\end{figure*}

We repeated the analysis with $3$--$4$ days fixed binning for two sources that have the best photon statistics in our sample, namely TXS~0506+056 and PKS~1502+106. We show the light curve for these sources with fixed binning overplotted on their adaptively-binned light curve in Appendix \ref{sec:appendixc} (see Fig \ref{fig:comp_adapt_fixed_txs}. and Fig \ref{fig:comp_adapt_fixed_pks}., respectively). The shapes of the light curves with fixed binning is similar to the adaptively-binned light curves, but with fixed binning the light curves are more noisy when the sources were faint. In comparing the results from the two light curve generation methods, we do not see different structures that would alter the results of this paper. Moreover, fixed binning of the light curve of the  other sources in our sample, being fainter than the two shown here, would lead to a large amount of non-detections. This further justifies the choice of the adaptive binning method employed in our work.

\begin{table*}
\caption{Characteristic timing and flux values of the minimum and maximum gamma-ray flux bins compared to the neutrino bins, and the $p$-values of the gamma-ray photon flux of the neutrino bin being the maximum flux in the light curve. The columns are: (1) source name, (2) number of gamma-ray bins of the light curve of the source ($N$), (3) MJD of the neutrino bin ($t_{F\nu}$), (4) average photon flux of the neutrino bin ($F_\nu$), (5) MJD of the minimum flux bin ($t_\mathrm{Fmin}$) minus MJD of the neutrino bin ($t_{F\nu}$), (6) average flux of the minimum flux bin ($F_\mathrm{min}$) over the flux of the neutrino bin ($F_\nu$), (7) MJD of the maximum flux bin ($t_\mathrm{Fmax}$) minus MJD of the neutrino bin ($t_{F\nu}$), (8) average flux of the maximum flux bin ($F_\mathrm{max}$) over the flux of the neutrino bin ($F_\nu$), (9) $p$-value ($p$).
\label{tab:pvalues}}
\centering
\begin{tabular}{ccccrcrcc}
\hline\hline
{Source} & $N$ & $t_{F\nu}$ & $F_\nu$ & $t_\mathrm{Fmin}-t_{F\nu}$ & $F_\mathrm{min}/F_\nu$ & $t_\mathrm{Fmax}-t_{F\nu}$ & $F_\mathrm{max}/F_\nu$ & $p$\\
(--) & (--) & (MJD) & ($\mathrm{ph}\,\mathrm{cm}^{-2}\,\mathrm{s}^{-1}$) & (days) & (--) & (days) & (--) & (--)\\
(1) & (2) & (3) & (4) & (5) & (6) & (7) & (8) & (9)\\
\hline
RBS 0958 & 8  &56417.0 & $(1.46 \pm 0.58)\times 10^{-8}$ & $123.2$ & $ 0.57 \pm 0.34$ & $ -113.8$ & $ 5.46 \pm 2.70$ & $0.519$\\
GB6 J1040+0617& 8  &56994.9 & $(1.57 \pm 0.46)\times 10^{-8}$ & $ -86.1$ & $  0.01 \pm 0.26$ & $ 0.0$ & $ 1.00 \pm 0.41$ & $0.133$\\
PKS 1313-333& 5  &57632.6 & $(5.94 \pm 2.44)\times 10^{-9}$ & $ 102.1$ & $ 0.39 \pm 0.37$ & $ -61.0$ & $ 4.36 \pm 2.34$ & $0.701$ \\
TXS 0506+056& 76  &58018.0 & $(1.10 \pm 0.32)\times 10^{-7}$ & $ -155.2$ & $ 0.20 \pm 0.11$ & $ -80.1$ & $ 2.94 \pm 0.99$ &$0.386$ \\
PKS 1454-354& 4  &58396.7 & $(2.73 \pm 1.80)\times 10^{-8}$ & $-98.30$ & $0.04 \pm 0.14$ & $169.03$ & $3.00 \pm 2.19$ & $0.372$\\ 
NVSS J042025-374443& 3  &58575.1 & $(9.50 \pm 4.93)\times 10^{-9}$ & $ 135.0 $ & $0.78 \pm 0.58$ & $ -105.0 $ & $1.43 \pm 0.92$ & $0.539$\\
PKS 0426-380& 49  &58607.8 & $(1.97 \pm 0.41)\times 10^{-7}$ & $ -143.6$ & $ 0.07 \pm 0.06$ & $ 43.5$ & $ 1.91 \pm 0.65$ & $0.219$\\
PKS 1502+106& 50  &58703.2 & $(5.40 \pm 1.01)\times 10^{-8}$ & $ 0.00$ & $ 1.00 \pm 0.26$ & $ -112.3$ & $ 6.31 \pm 1.91$ & $0.935$\\
\hline
\end{tabular}
\end{table*}
\section{Results}
\label{section:results}

Below we refer to the centers and width of the bins in the form of center \; $\pm$ \; width/2, to help understand the time-characteristics of the bins. We note here the $\pm$ does not mean an error. We modeled the light curves of the neutrino source candidates between $100$~MeV and $300$~GeV, except for GB6 J1040+0617, PKS 1313-333 and TXS 0506+056, where, to avoid the contamination by the neighboring bright gamma-ray sources, we set the minimum energy of the analysis to $300$~MeV. We fitted the normalization factor and the spectral index of the source of interest in each bin. Normalization of sources with TS-value$>25$ and sources within $5$~degrees of the source of interest were also free (regardless of their test statistic). To better understand significant variations in the light curves, we applied the Bayesian Block algorithm of \citet{Scargle2013} to them. Following \citet{Garrappa2019}, we set the probability to $p=0.05$. The Bayesian blocks are shown on the light curve of the gamma-ray sources in the Appendix as stated below from source-to-source. We summarize the characteristic timing and flux values of minimum and maximum photon flux bins compared to the neutrino bins in Table \ref{tab:pvalues}., and elaborate more on the results below.

\begin{figure}[h!]
\centering
\includegraphics[width=0.85\textwidth]{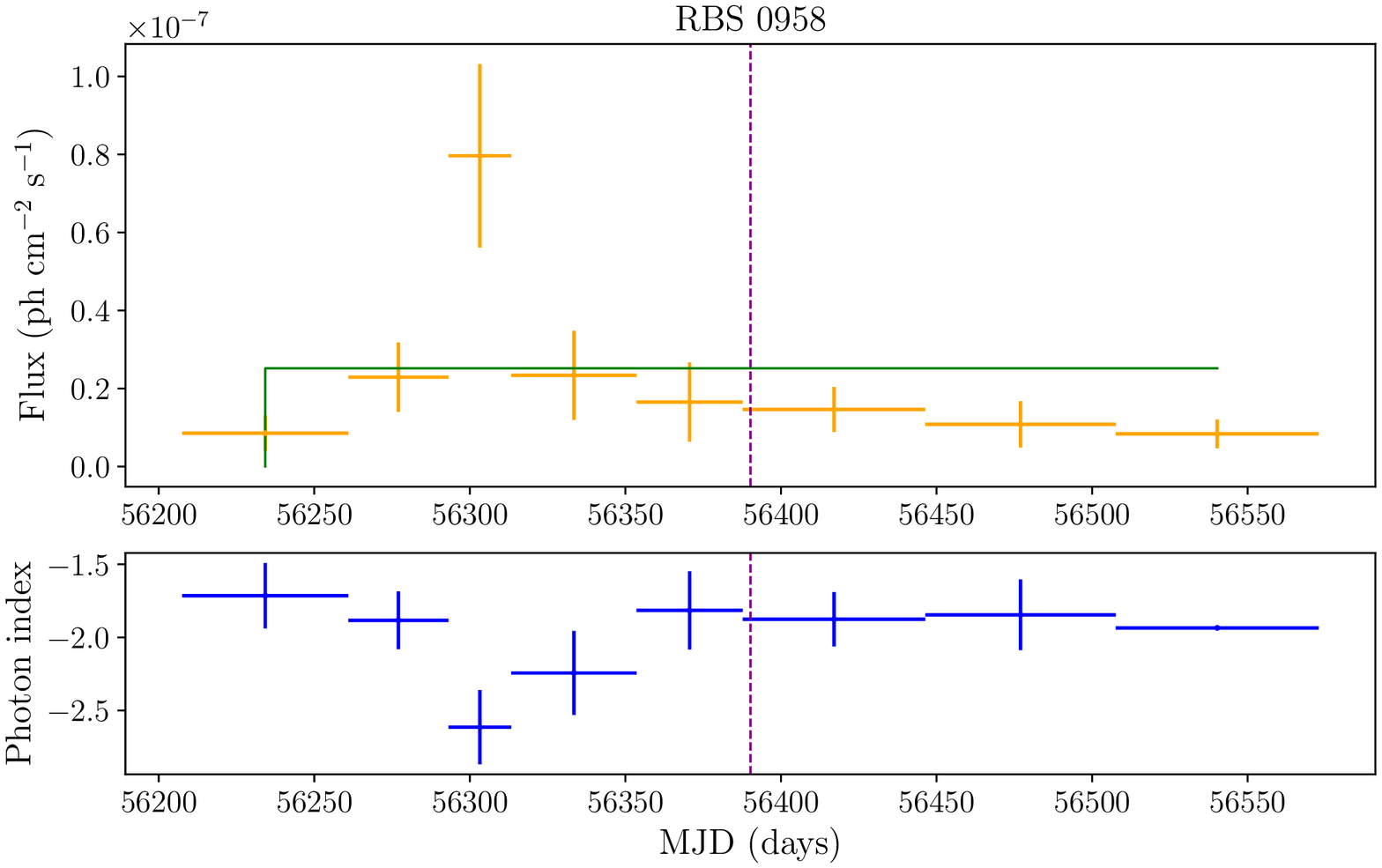}
\caption{Upper panel: Likelihood light curve of RBS 0958 with Bayesian blocks ($p=0.05$). The green continuous line represents the Bayesian blocks binning. Lower panel: Photon index as function of time.}
\label{fig:rbs}
\end{figure}

\textbf{RBS~0958} is located $1.19\degr$ away from the best-fit localization of the neutrino event IC-130408A, within the 90\% error ellipse of the neutrino. We show the one-year long likelihood light curve of RBS~0958 in Fig \ref{fig:rbs}. along with the photon index as function of time. The corresponding neutrino was detected in $T_\nu=\mathrm{MJD}~56390.1887627$, with reconstructed neutrino energy of $E_\nu=30.8^{+3.3}_{-3.5}$~TeV. The neutrino came in the \textit{Fermi}-LAT bin MJD $56417.0\pm29.4$ days, with an average flux of $(1.46\pm  0.58) \times 10^{-8}~\mathrm{ph}\,\mathrm{cm}^{-2}\,\mathrm{s}^{-1}$. This bin is situated on the decreasing branch of a gamma flare, which is centered to MJD $56303.2\pm10.0$ days ($86.97$ days before the neutrino-detection). Though the Bayesian blocks (with $p=0.05$) do not indicate a significantly variable flux during the one-year long time window (see Fig \ref{fig:rbs}.), the peak photon flux of the suspected flare ($(8.0\pm 2.4)\times 10^{-8} ~\mathrm{ph}\,\mathrm{cm}^{-2}\,\mathrm{s}^{-1}$) is about 45--83 times higher than the average photon flux of RBS~0958 in the 4FGL-DR2 catalog ($1.24\times10^{-9}~\mathrm{ph}\,\mathrm{cm}^{-2}\,\mathrm{s}^{-1}$). This indicates that there might indeed be a gamma-ray flare going on in this window. Based on its photon indices, RBS~0958 shows the sign of spectral softening during the gamma-ray flare. The peak-flux of this flare was $(7.97\pm2.35) \times 10^{-8}~\mathrm{ph}\,\mathrm{cm}^{-2}\,\mathrm{s}^{-1}$, which is about $5.5$ times the flux of the bin in the gamma-ray light curve containing the detected neutrino (from here we refer to this specific bin of the gamma-ray light curve as "neutrino bin"). The bins before (MJD~$56370.6\pm17.0$) and after (MJD~$56477.0\pm30.6$) the neutrino-bin have similar fluxes than the flux of the neutrino-bin ($(1.65\pm1.02) \times 10^{-8}~\mathrm{ph}\,\mathrm{cm}^{-2}\,\mathrm{s}^{-1}$ and $(1.08\pm0.59) \times 10^{-8}~\mathrm{ph}\,\mathrm{cm}^{-2}\,\mathrm{s}^{-1}$, respectively).  

\begin{figure*}
\centering
\includegraphics[width=0.85\textwidth]{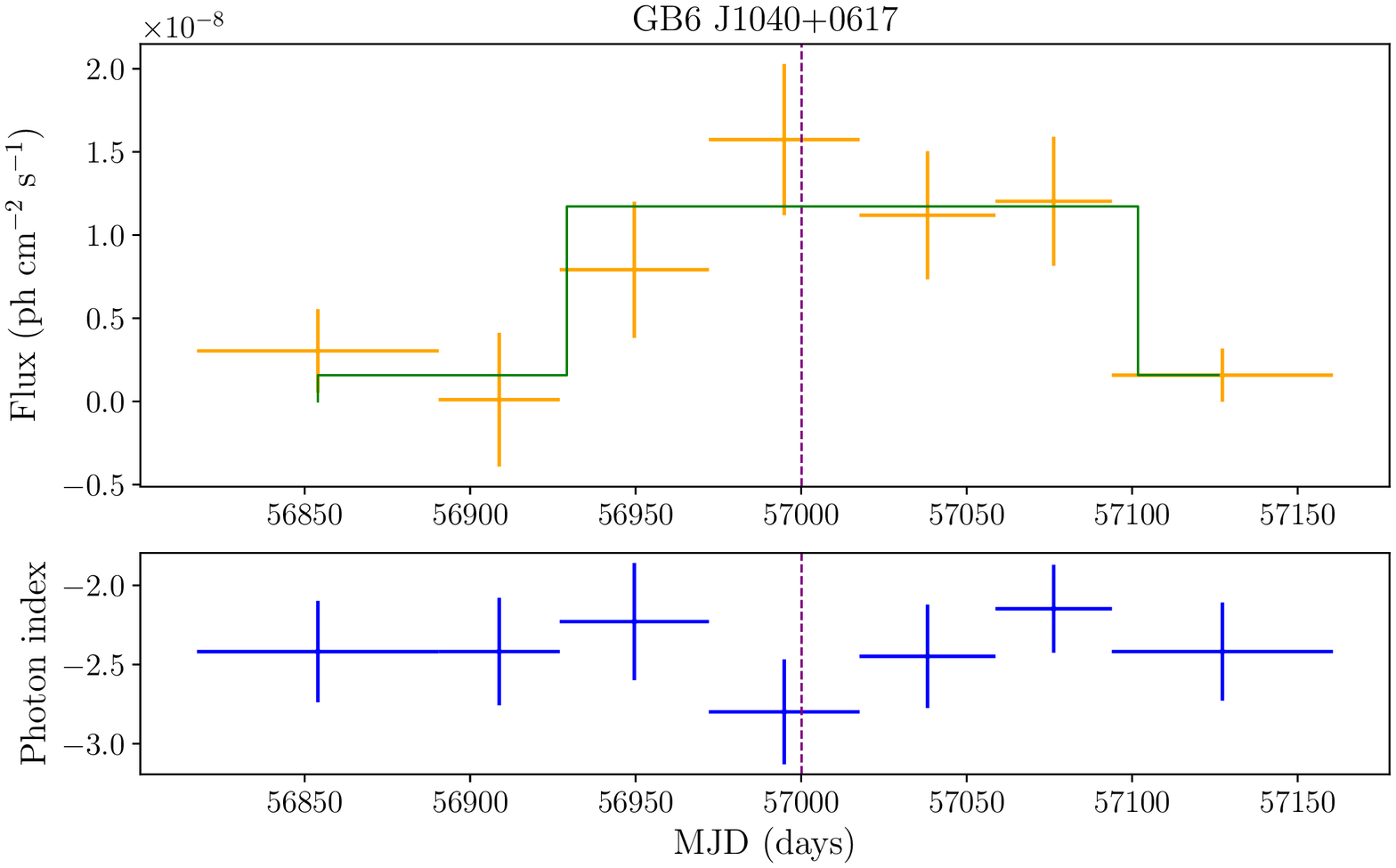}
\caption{Upper panel: Likelihood light curve of GB6 J1040+0617 with Bayesian blocks ($p=0.05$). The green continuous line represents the Bayesian blocks binning. Lower panel: Photon index as function of time.}
\label{fig:gb6}
\end{figure*}

\textbf{GB6~J1040+0617} is located within the $90$\% error ellipse of IC-141209A, $0.45\degr$ away from the best-fit localization of the neutrino. We show the one-year long likelihood light curve of GB6 J1040+0617 in Fig \ref{fig:gb6} along with the photon index as function of time. The corresponding neutrino was detected in $T_\nu=\mathrm{MJD}~57000.14311$, with reconstructed neutrino energy of $E_\nu=97.4^{+9.6}_{-9.6}$ TeV. The neutrino came in the \textit{Fermi}-LAT bin MJD $56994.9\pm22.8$ days, with an average flux of $(1.57\pm 0.46) \times 10^{-8}~\mathrm{ph}\,\mathrm{cm}^{-2}\,\mathrm{s}^{-1}$. This is the peak flux of the gamma-ray flare shown by this source during the one year long light curve centered to the time of the neutrino detection. The flux in the previous bin (MJD~$56949.6\pm22.5$) is about half the flux of the neutrino-bin ($(7.9\pm4.1) \times 10^{-9}~\mathrm{ph}\,\mathrm{cm}^{-2}\,\mathrm{s}^{-1}$). The flux in the next bin (MJD~$57038.2\pm20.5$) is about $2/3$ of the flux in the neutrino-bin ($(1.11\pm0.38) \times 10^{-8}~\mathrm{ph}\,\mathrm{cm}^{-2}\,\mathrm{s}^{-1}$). 

\begin{figure*}
\centering
\includegraphics[width=0.85\textwidth]{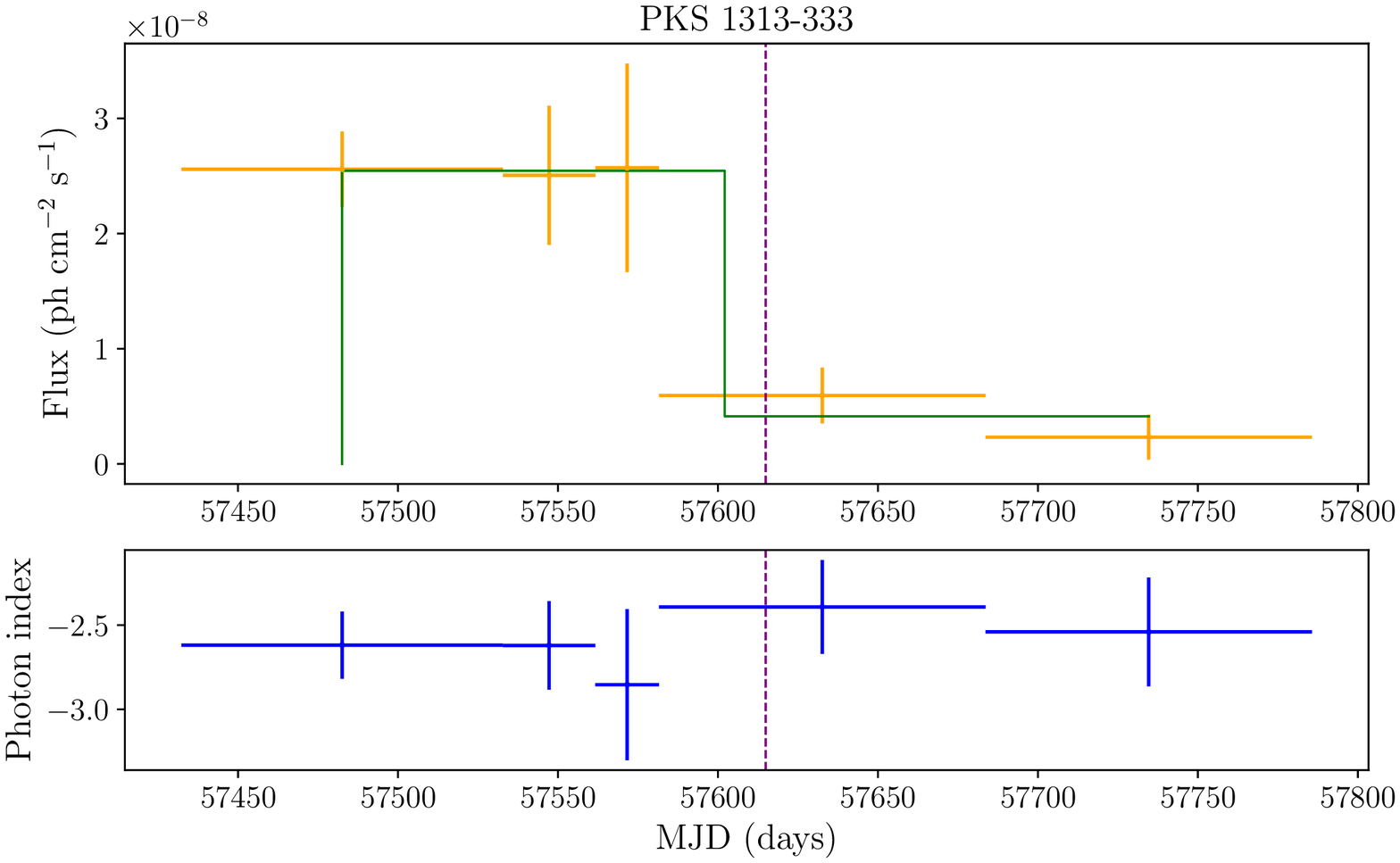}
\caption{Upper panel: Likelihood light curve of PKS 1313-333 with Bayesian blocks ($p=0.05$). The green continuous line represents the Bayesian blocks binning. Lower panel: Photon index as function of time.}
\label{fig:pks1313}
\end{figure*}

\textbf{PKS~1313-333} is located within the 90\% error ellipse of IC-160814A ($T_\nu=\mathrm{MJD}~57614.91$), $1.29\degr$ away from its best-fit sky coordinates. We show the one-year long likelihood light curve of PKS~1313-333 in Fig \ref{fig:pks1313} along with the photon index as function of time. The neutrino came in the \textit{Fermi}-LAT bin MJD~$57632.6\pm51.0$, with flux of $(5.9\pm 2.4) \times 10^{-9}~\mathrm{ph}\,\mathrm{cm}^{-2}\,\mathrm{s}^{-1}$. The previous bin, which is the peak of the light curve, is situated at MJD~$57571.6\pm10.0$ with flux of $(2.57\pm0.91) \times 10^{-8}~\mathrm{ph}\,\mathrm{cm}^{-2}\,\mathrm{s}^{-1}$, which is more then 4 times the flux of the neutrino bin. The next bin (MJD~$57734.7\pm51.0$) has an average flux of $(2.3\pm2.0) \times 10^{-9}~\mathrm{ph}\,\mathrm{cm}^{-2}\,\mathrm{s}^{-1}$, about half of the flux of the neutrino bin. This is also the faintest bin of the light curve.

\begin{figure*}
\centering
\includegraphics[width=0.85\textwidth]{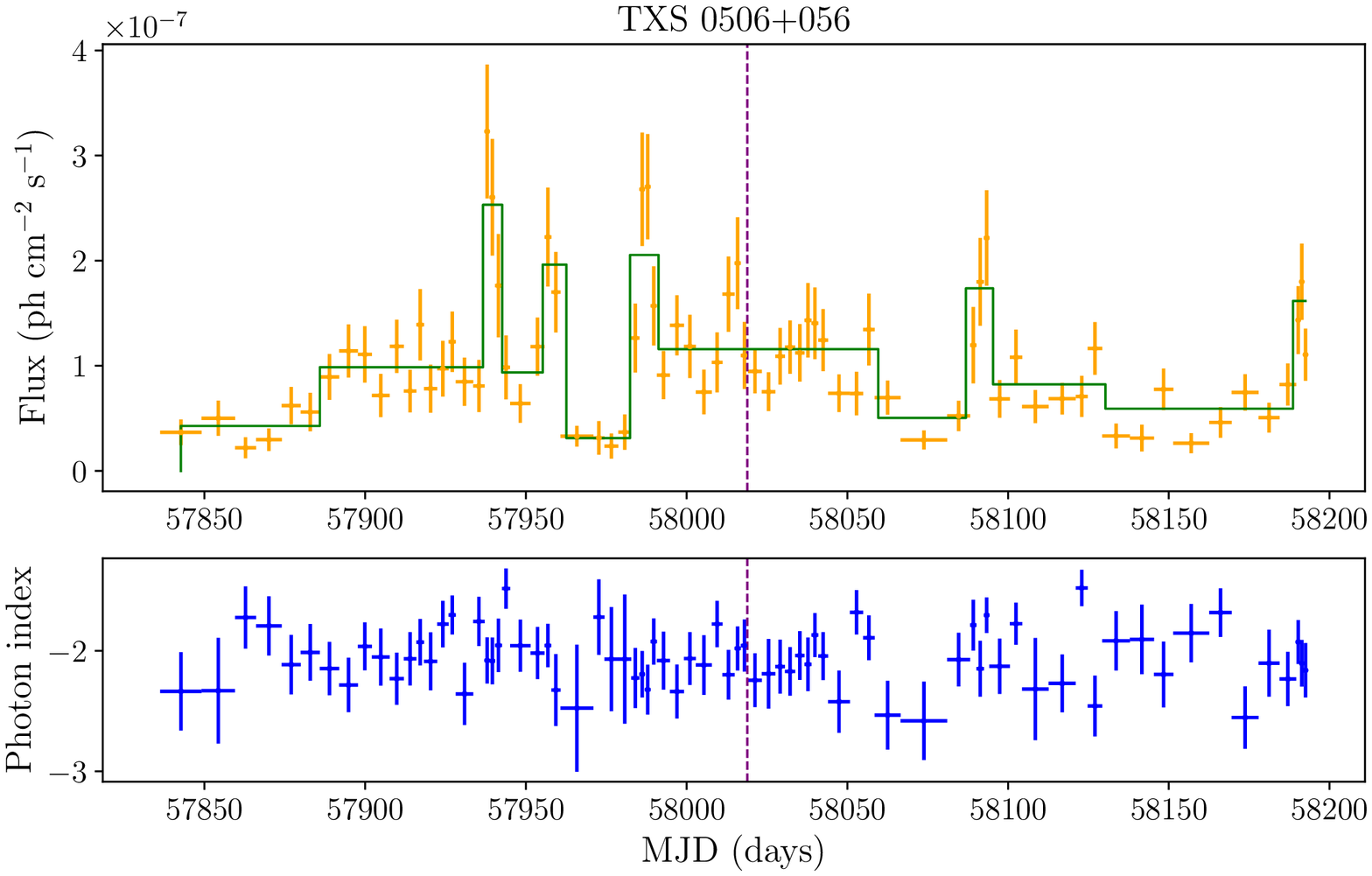}
\caption{Upper panel: Likelihood light curve of TXS 0506+056 with Bayesian blocks ($p=0.05$). The green continuous line represents the Bayesian blocks binning. Lower panel: Photon index as function of time.}
\label{fig:txs}
\end{figure*}

\textbf{TXS~0506+056} is located within the $90$\% error ellipse of IC-170922A ($T_\nu=\mathrm{MJD}~58018.87$), $0.1\degr$ away from its best-fit sky coordinates. We show the one-year long likelihood light curve of TXS~0506+056 in Fig \ref{fig:txs} along with the photon index as function of time. The neutrino came in the $2.30$ days long \textit{Fermi}-LAT bin centered to MJD~$58018.0$, that has a the flux of $(1.10\pm 0.32) \times 10^{-7}~\mathrm{ph}\,\mathrm{cm}^{-2}\,\mathrm{s}^{-1}$. This bin is on the decreasing side of a gamma-ray dip. The highest-flux bin (MJD~$57938.0\pm0.8$) has a flux of $(3.23\pm0.64) \times 10^{-7}~\mathrm{ph}\,\mathrm{cm}^{-2}\,\mathrm{s}^{-1}$, which is about three times the flux in the neutrino bin. The highest-flux bin is situated $80.0$ days before the detection time of the neutrino. The minimum flux of the light curve  $(2.2\pm0.1) \times 10^{-8}~\mathrm{ph}\,\mathrm{cm}^{-2}\,\mathrm{s}^{-1}$ is reached in the bin MJD $57962.8\pm3.3$, centered $155$ days prior to the detection.

\begin{figure*}
\centering
\includegraphics[width=0.85\textwidth]{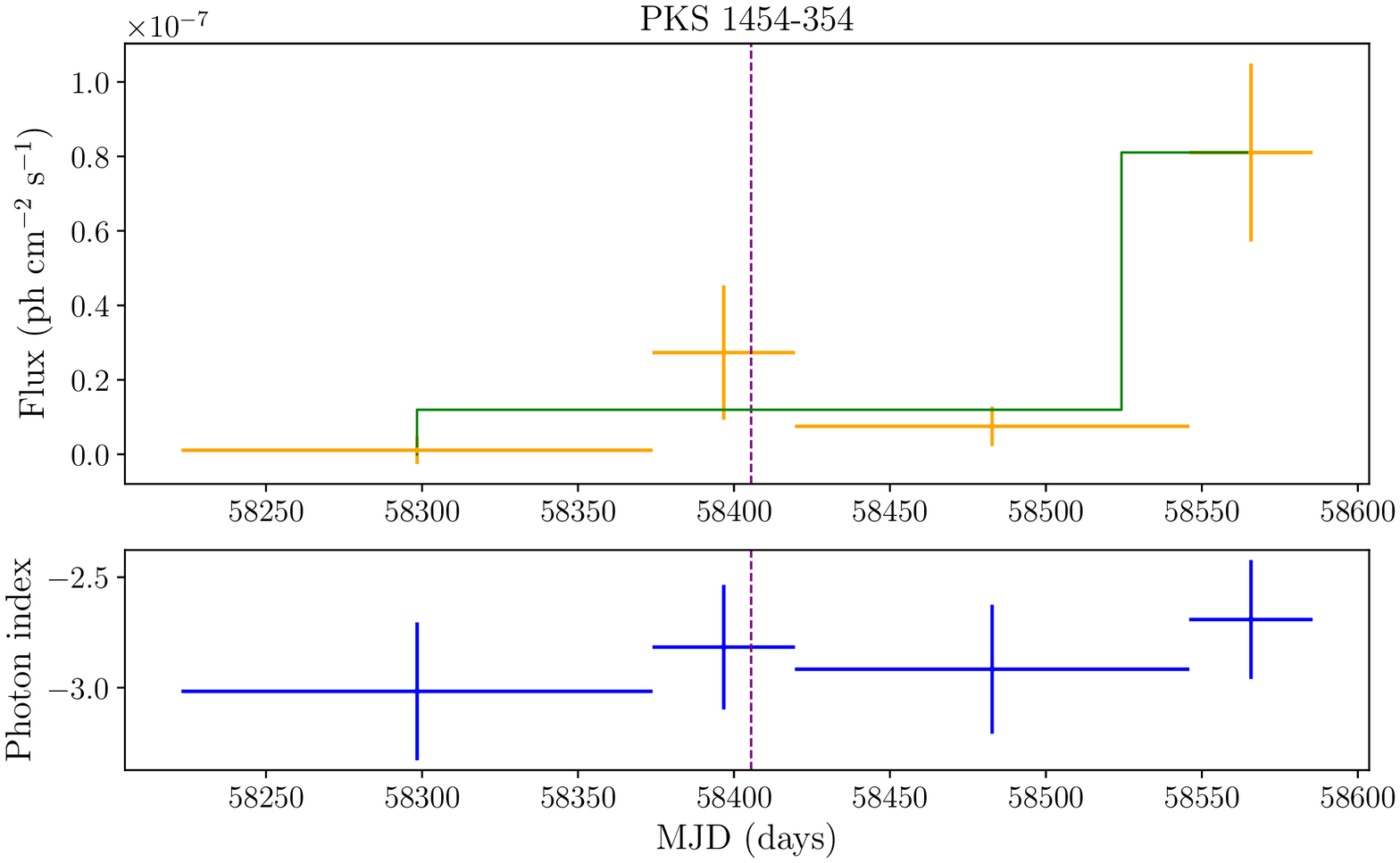}
\caption{Upper panel: Likelihood light curve of PKS 1454-354 with Bayesian blocks ($p=0.05$). The green continuous line represents the Bayesian blocks binning. Lower panel: Photon index as function of time.}
\label{fig:pks1454}
\end{figure*}

\textbf{PKS~1454-354} is located within the 90\% error ellipse of IC-181014A ($T_\nu=\mathrm{MJD}~58405.5$), $0.85\degr$ away from best-fit sky coordinates of this neutrino event. We show the one-year long likelihood light curve of PKS~1454-354 in Fig \ref{fig:pks1454} along with the photon index as function of time. The neutrino came in the \textit{Fermi}-LAT bin MJD~$58396.7\pm22.8$, with the photon flux of $(2.7\pm 1.8) \times 10^{-8}~\mathrm{ph}\,\mathrm{cm}^{-2}\,\mathrm{s}^{-1}$, indicating the source was in a faint phase.  We note that the photon flux in the neutrino bin is about an order of magnitude larger than the average photon flux of this source in the 4FGL-DR2 catalog ($3.49\times 10^{-9}~\mathrm{ph}\,\mathrm{cm}^{-2}\,\mathrm{s}^{-1}$), and that it is about an order of magnitude smaller than the peak photon flux of this source in the same catalog ($2.56\times 10^{-7}~\mathrm{ph}\,\mathrm{cm}^{-2}\,\mathrm{s}^{-1}$).

\begin{figure*}
\centering
\includegraphics[width=0.85\textwidth]{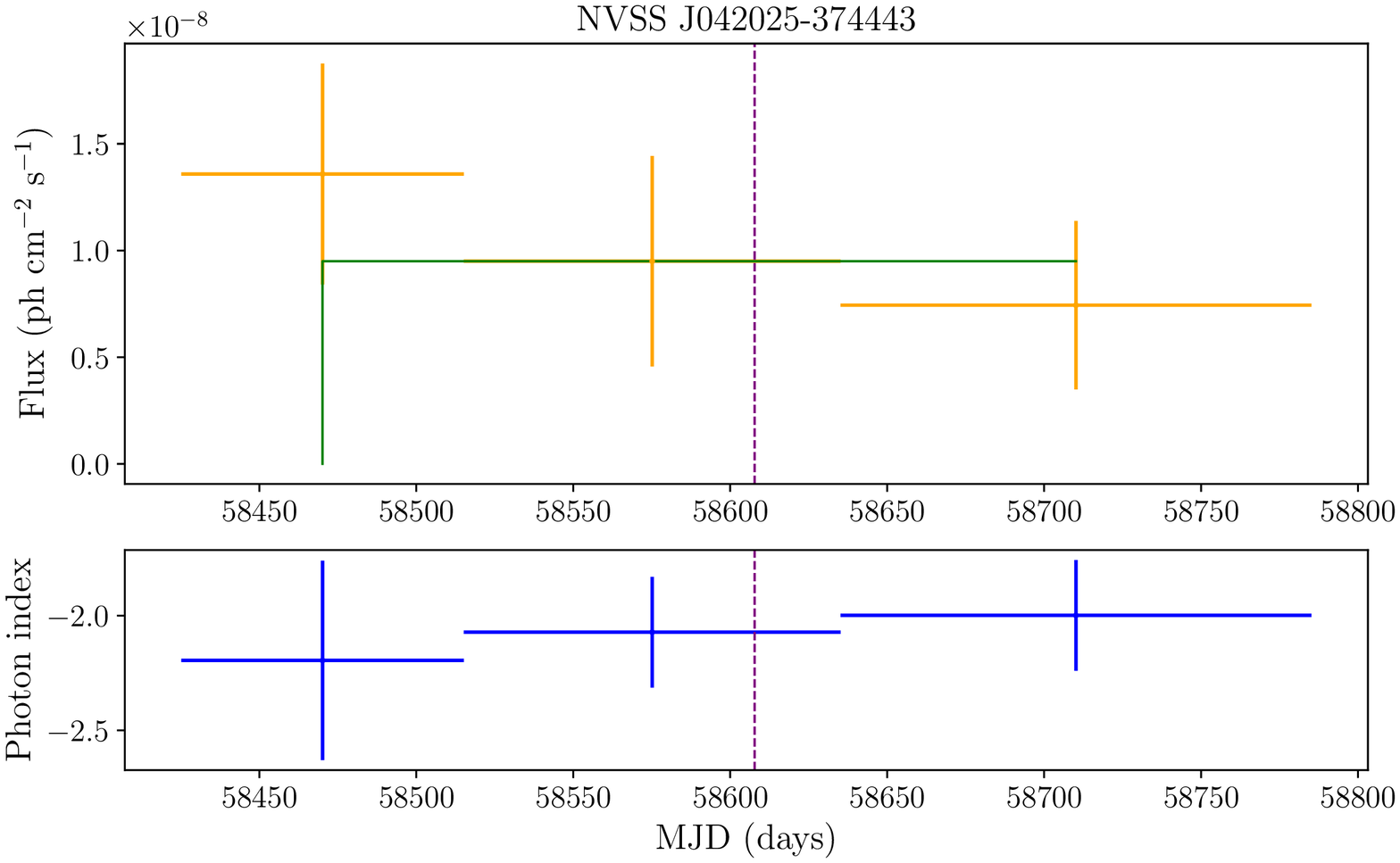}
\caption{Upper panel: Likelihood light curve of NVSS J042025-374443 with Bayesian blocks ($p=0.05$). The green continuous line represents the Bayesian blocks binning. Lower panel: Photon index as function of time.}
\label{fig:nvss}
\end{figure*}

\textbf{NVSS~J042025-374443} is located within the $90$\% error ellipse of IC-190504A ($T_\nu=\mathrm{MJD}~58607.77$), $0.76\degr$ away from its best-fit sky coordinates. We show the one-year long likelihood light curve of NVSS~J042025-374443 in Fig \ref{fig:nvss} along with the photon index as function of time. Though the source is rather faint in the studied observational window and therefore we were able to get only 3 bins of the light curve, we can note the followings. The neutrino came in the \textit{Fermi}-LAT bin MJD~$58575.1\pm60$, with flux of $(9.5\pm4.9) \times 10^{-9}~\mathrm{ph}\,\mathrm{cm}^{-2}\,\mathrm{s}^{-1}$. The previous bin is situated as MJD~$58470.1\pm45.0$, with flux of $(1.36\pm0.52) \times 10^{-8}~\mathrm{ph}\,\mathrm{cm}^{-2}\,\mathrm{s}^{-1}$. The next bin MJD~$58710.1\pm75$ has an average flux of $(7.4\pm3.9) \times 10^{-9}~\mathrm{ph}\,\mathrm{cm}^{-2}\,\mathrm{s}^{-1}$. The highest-flux bin MJD~$58470.1\pm45.0$ is centered $137.63$ days before the detection time of the corresponding neutrino, and it has an average flux of $(1.36\pm0.52) \times 10^{-8}~\mathrm{ph}\,\mathrm{cm}^{-2}\,\mathrm{s}^{-1}$. We note that the photon flux of the neutrino bin is $4$--$13$ times higher than the average photon flux of this source in the 4FGL-DR2 catalog ($1.09\times 10^{-9}~\mathrm{ph}\,\mathrm{cm}^{-2}\,\mathrm{s}^{-1}$), and it is about $2$--$8$ times smaller compared to the peak photon flux of the source in the 4FGL-DR2 catalog ($3.56\times 10^{-8}~\mathrm{ph}\,\mathrm{cm}^{-2}\,\mathrm{s}^{-1}$).

\begin{figure*}
\centering
\includegraphics[width=0.85\textwidth]{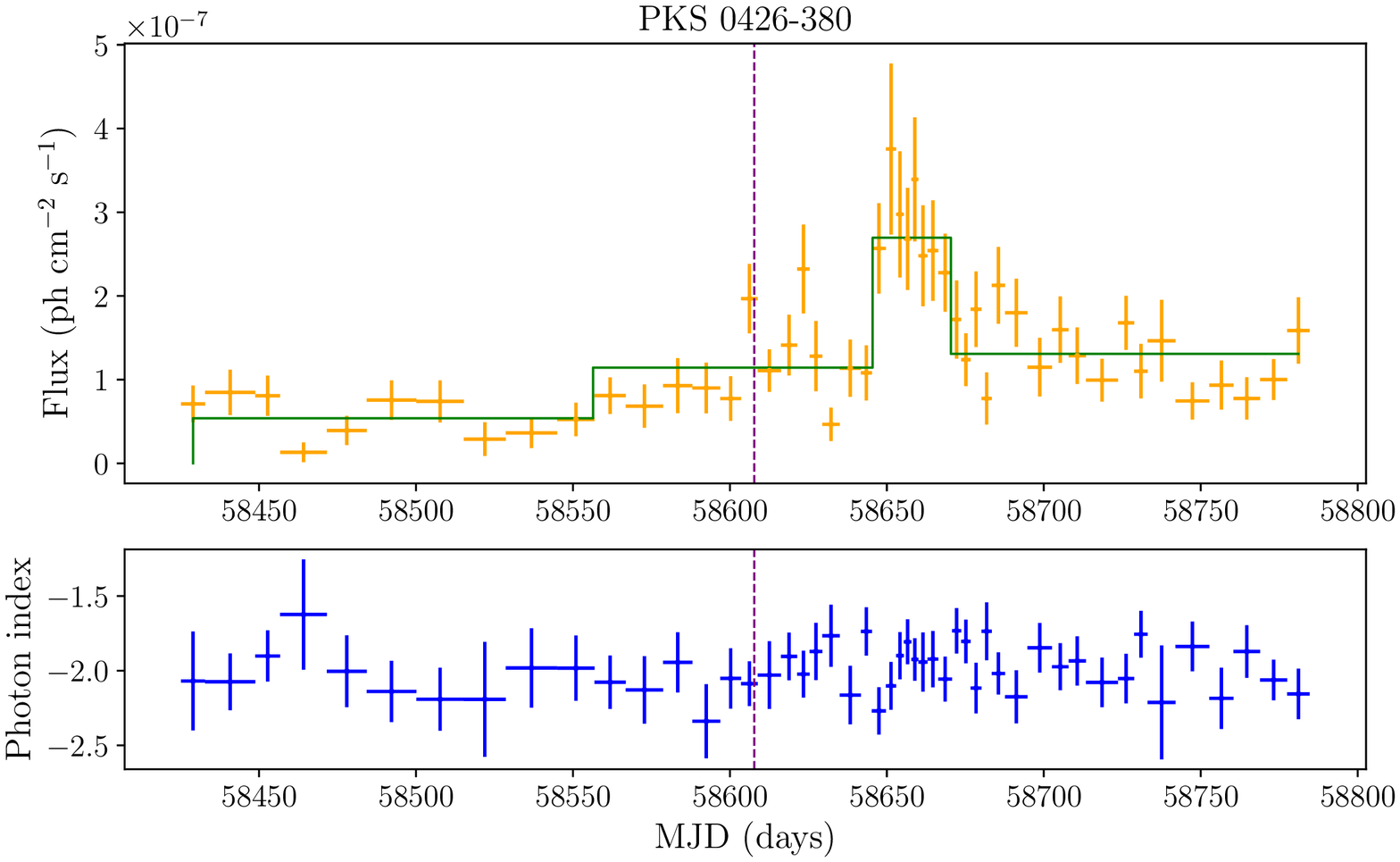}
\caption{Upper panel: Likelihood light curve of PKS 0426-380 with Bayesian blocks ($p=0.05$). The green continuous line represents the Bayesian blocks binning. Lower panel: Photon index as function of time.}
\label{fig:pks0426}
\end{figure*}

\textbf{PKS~0426-380} is also located within the 90\% error ellipse of IC-190504A ($T_\nu=\mathrm{MJD}~58607.77$), $1.47 \degr$ away from its best-fit sky positions. We show the one-year long likelihood light curve of PKS~0426-380 in Fig \ref{fig:pks0426} along with the photon index as function of time. The neutrino came in the \textit{Fermi}-LAT bin MJD~$58607.8\pm2.7$, that has an average flux of $(1.97\pm0.41) \times 10^{-7}~\mathrm{ph}\,\mathrm{cm}^{-2}\,\mathrm{s}^{-1}$. The highest-flux gamma-ray bin MJD~$58651.3\pm1.6$ is centered $43.6$ days after the neutrino, and the peak of a gamma-ray flare shown by the source. This light curve bin has an average flux of $(3.76\pm1.02) \times 10^{-7}~\mathrm{ph}\,\mathrm{cm}^{-2}\,\mathrm{s}^{-1}$, which is about 2 times higher compared to the flux of the neutrino bin. PKS~0426-380 is a significantly stronger gamma source than the other gamma source, NVSS~J042025-374443, that also lies within the $90$\% error ellipse of IC-190504A.

\begin{figure*}
\centering
\includegraphics[width=0.85\textwidth]{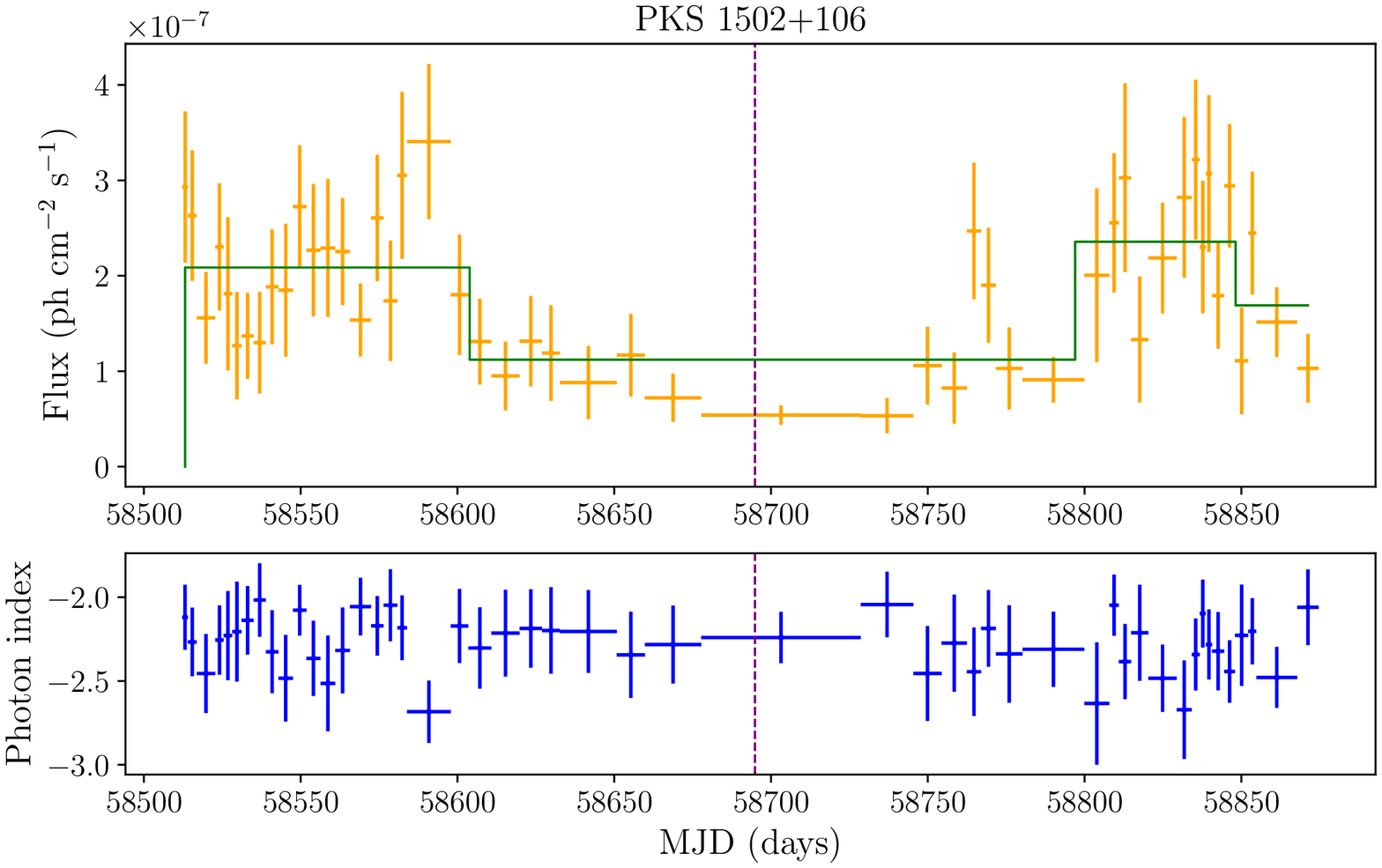}
\caption{Upper panel: Likelihood light curve of PKS 1502+106 with Bayesian blocks ($p=0.05$). The green continuous line represents the Bayesian blocks binning. Lower panel: Photon index as function of time.}
\label{fig:pks1502}
\end{figure*}

\textbf{PKS~1502+106} is located within the $90$\% error ellipse of IC-190730A ($T_\nu=\mathrm{MJD}~58694.87$), $0.58\degr$ away from the best-fit sky coordinates of the corresponding neutrino. This $299$~TeV event was distributed through the Gold channel of the IceCube Real-Time Alert System. Its signalness is $0.67$, and the false alarm rate is $0.677$ $yr^{-1}$. We show the one-year likelihood light curve of PKS~1502+106 in Fig \ref{fig:pks1502} along with the photon index as function of time. The neutrino came in the \textit{Fermi}-LAT bin MJD $58703.2\pm25.5$, that has a flux of $(5.4\pm1.0) \times 10^{-8}~\mathrm{ph}\,\mathrm{cm}^{-2}\,\mathrm{s}^{-1}$. This bin marks the global minimum of the one-year light curve of PKS~1502-106, the lowest flux bin of a prolonged deep minimum lasting about $178$ days. The highest-flux bin (MJD~$58590.9\pm7.0$), centered $112$ days before the neutrino detection, has an average flux of $(3.4\pm0.8) \times 10^{-7}~\mathrm{ph}\,\mathrm{cm}^{-2}\,\mathrm{s}^{-1}$) which is more than $6$ times the flux of the neutrino bin.

\section{Discussion and conclusions}
\label{section:disc}

Low optical depth high-energy neutrino formation processes predict that the neutrino flux is proportional to the gamma-ray photon flux and one therefore expects the emission of high-energy neutrinos during gamma flares.

We calculated the chance probability of detecting a neutrino in a bin with a gamma-ray photon flux higher then the observed gamma-ray photon flux of the neutrino bin:
\begin{equation}
p=P(F_{\gamma,max}>F_{\gamma,\nu})=\frac{\Sigma_i \int^{\infty}_{F_{\gamma,\nu}}  \mathcal{N}(x,F_i,\sigma_i) dFx}{\Sigma_i \int^{\infty}_{-\infty}  \mathcal{N}(x,F_i,\sigma_i) dFx}
\end{equation}
where $\mathcal{N}(x,F_i,\sigma_i)$ is the Gaussian evaluated at $x$, when we assume the uncertainty of the flux is normally distributed. When the p-value is close to zero, it means the neutrino came in gamma maximum, and there is no chance of getting higher gamma-ray photon flux during the studied time range. This is the case when the medium is transparent to the gamma photons and we see a gamma peak during the detection of high-energy neutrinos from the source.

When the $p$-value is close to unity, it means the neutrino came in a gamma minimum and therefore all bins of the light curve are brighter than the flux of the bin in which the neutrino arrived. This is the case when the neutrino flux and gamma-ray photon flux are anti-correlated e.g. due to the high in-source gamma-gamma optical depth, and at the time of maximum neutrino emissivity the pionic gamma photons are maximally 
absorbed and cascade down to X-ray energies. This is the case of beam dampening. Using the above considerations we calculated the p-values (see in Table \ref{tab:pvalues}).

It seems the corresponding neutrino came in the global maximum of the one-year long gamma light curve of GB6~J1040+0617, and in global minimum of the one-year long gamma light curve of PKS~1502+106. For the other neutrino source candidates, PKS~1313-333, PKS~1454-354 NVSS~J042025-374443, RBS~0958, TXS~0506+056, PKS~0426-380, the neutrino arrival time seems to not correlate with the gamma-ray photon flux, even though fainter phases are more characteristic to the neutrino arrivals.

What we can conclude is that apart from GB6~J1040+0617, none of the neutrino source candidates were significantly bright in a time window of $\pm6$ months around the corresponding IceCube neutrinos. We note here that our findings are based on a one-year long gamma-ray photon flux curve of the respective sources. Taking into account the whole \textit{Fermi}-LAT light curves might alter the outcomes of this analysis. 

The fact that, out of the $8$ neutrino candidate sources (associated with $7$ neutrino events) studied, we found only one neutrino source candidate blazar at peak gamma-ray photon flux during the neutrino emission means that either the sources in our sample are not the sources of the corresponding IceCube neutrinos, or an in-source effect \citep[e.g. suppression of gamma rays due to high gamma-gamma opacity,][]{Kun2021} complicates the multimessenger scenario of neutrino emission at least for these blazars. It is also possible that a mix of these two scenarios is in play. Future studies conducted at even higher energies for which gamma-ray absorption should be more evident or alternative measurements, such as HI column density measurements, might be helpful to disambiguate between these scenarios.

\begin{acknowledgements}
We thank Simone Garrappa, Deirdre Horan, Josefa Becerra Gonzalez, Melissa Pesce Rollins and Patrik Mil{\'a}n Veres for comments and discussions. We acknowledge support from the Deutsche Forschungsgemeinschaft DFG, within the Collaborative Research Center SFB1491 "Cosmic Interacting Matters - From Source to Signal" (project No. 445052434). E.K. thanks the Alexander von Humboldt Foundation for its Postdoctoral Fellowship. I.B. acknowledges the support of the Alfred P. Sloan Foundation and NSF grants PHY-1911796 and PHY-2110060. This paper makes use of publicly available Fermi-LAT data provided online by the \url{https://fermi.gsfc.nasa.gov/ssc/data/access/} Fermi Science Support Center. On behalf of Project 'fermi-agn' we thank for the usage of the ELKH Cloud that significantly helped us achieving the results published in this paper. 
\end{acknowledgements} 


\begin{appendix}
\onecolumn
\section{Appendix A: Additional sources found in the ROI of the target sources}

\begin{table}[h!]
\centering
\caption{Additional sources in the ROI about the target sources. Source ID (1), right-ascension J2000 (2), declination J2000 (3), offset of the new source to the target source (4),  flux measured by Fermi between $0.1$\,GeV and $100$\,GeV (5), error of the flux value (6), TS-value of the detection (7), predicted number of photons (8).\label{table:addsources}}
\begin{tabular}{cccccccc}
\hline
\hline
{Source ID} &{RA} & {DEC} & {$\Delta d$} & {$F_{1000}\times 10^{10}$} & {$\mathrm{err}F_{1000}\times 10^{10}$} & {TS} & {$N_\mathrm{pred}$}\\
{-} & {$(\degr)$} & {$(\degr)$} & {$(\degr)$} & {$(\mathrm{ph}~\mathrm{cm}^{-2}~\mathrm{s}^{-1}$)} & {$(\mathrm{ph}~\mathrm{cm}^{-2}~\mathrm{s}^{-1})$} & {-} & {-}\\
(1) &(2) &(3) &(4) &(5) &(6) &(7) &(8)\\
\hline
\multicolumn{8}{c}{RBS 0958}\\
PS J1106.9+2708 &166.719 &27.139& 7.29 &2.39 &1.13& 18.38& 154.21\\
PS J1132.6+2738 & 173.161 & 27.642 & 8.219 & 7.80 & 1.97 & 58.02 & 60.53\\
\hline
\multicolumn{8}{c}{GB6 J1040+0617}\\
PS J1042.6+0838& 160.655 &8.637 &2.405& 7.89& 1.72 &441.27 &615.38\\
\hline
\multicolumn{8}{c}{PKS 1313-333}\\
PS J1242.8-2554& 190.720& -25.912& 10.557 & 5.07 &1.70 & 8.1 &32.42\\
PS J1246.4-2602& 191.610& -26.039& 9.950& 8.68 &0.12& 60.65 &123.66\\
\hline
\multicolumn{8}{c}{TXS 0506+056}\\
PS J0500.7-0140 &75.183& -1.671& 7.686& 22.32 &0.18& 365.95 &383.90\\
\hline
\multicolumn{8}{c}{NVSS J042025-374443}\\
PS J0347.8-3618 &56.952 &-36.316 &6.654 &4.62 &1.25 &40.84  &147.55\\
PS J0348.7-3029 &57.184 &-30.484 &9.774 &3.06 &1.01 &131.58 &424.67\\
PS J0454.4-3140 &73.614 &-31.667 &9.271 &2.44 &1.09 &20.71 &133.68\\
\hline
\multicolumn{8}{c}{PKS 0426-380}\\
PS J0456.4-4318 &74.113 &-43.302 &7.511 &3.80 &1.31 &53.98 &370.24\\
\hline
\multicolumn{8}{c}{PKS 1502+106}\\
PS J1446.6+1213 &221.652 &12.216 &4.690 &3.37 &0.14 &16.96 &60.75\\
PS J1525.8+1636 &231.470 &16.605 &8.030 &4.14 &0.33 &48.66 &350.34\\
\hline
\end{tabular}
\end{table}

\section{Appendix B: Light curves}
\vspace{0.5cm}

\vspace{0.5cm}
    
\vspace{0.5cm}

\clearpage
\section{Appendix C: Comparison of adaptive and fixed binning of the light curve of two bright sources}
\label{sec:appendixc}

\begin{figure*}
\centering
\includegraphics[width=0.4\textwidth,angle=270]{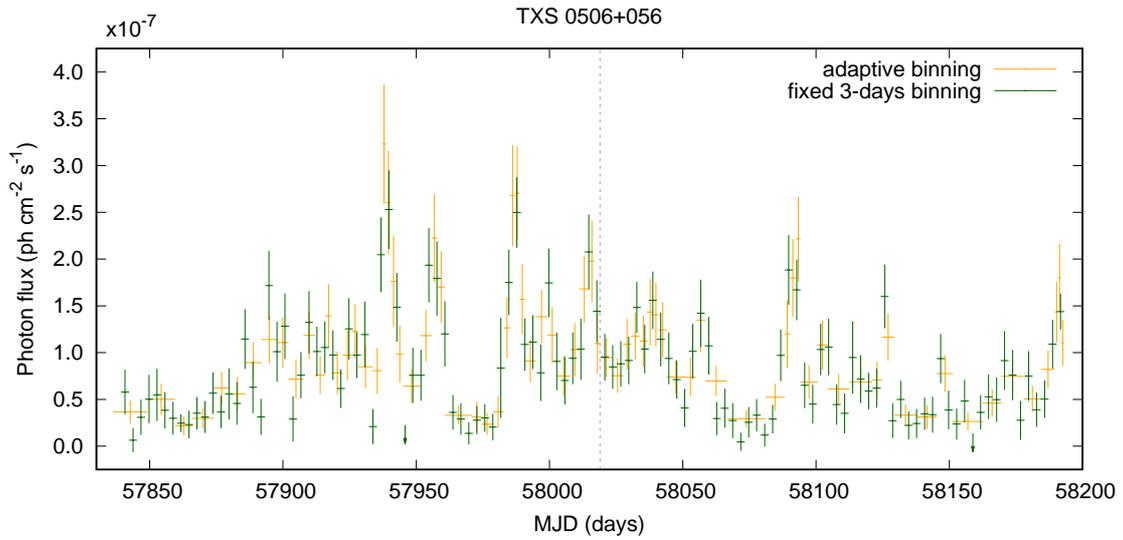}
\caption{This plot shows the light curve of TXS~0506+056 with fixed binning (green crosses), overplotted on its adaptively-binned light curve (orange crosses). For very faint bins of the fixed binning light curve we plot only upper limits (down-pointing green arrows).}
\label{fig:comp_adapt_fixed_txs}
\end{figure*}

\end{appendix}
\end{document}